# Electrical Investigation of the Oblique Hanle Effect in Ferromagnet/Oxide/Semiconductor Contacts

Kun-Rok Jeon[1], Byoung-Chul Min[2], Youn-Ho Park[2], Seung-Young Park[3], and Sung-Chul Shin[1,4*]

[1]*Department of Physics and Center for Nanospinics of Spintronic Materials, Korea Advanced Institute of Science and Technology (KAIST), Daejeon 305-701, Korea*

[2]*Center for Spintronics Research, Korea Institute of Science and Technology (KIST), Seoul 136-791, Korea*

[3]*Nano Materials Research Team, Korea Basic Science Institute (KBSI), Daejeon 305-764, Korea*

[4]*Department of Emerging Materials Science, Daegu Gyeongbuk Institute of Science and Technology (DGIST), Daegu 711-873, Korea*

[*]e-mail: scshin@kaist.ac.kr; scshin@dgist.ac.kr

**We have investigated the electrical Hanle effect with magnetic fields applied at an oblique angle ($\theta$) to the spin direction (the oblique Hanle effect, OHE) in CoFe/MgO/semiconductor (SC) contacts by employing a three-terminal measurement scheme. The electrical oblique Hanle signals obtained in CoFe/MgO/Si and CoFe/MgO/Ge contacts show clearly different line shapes depending on the spin lifetime of the host SC. Notably, at moderate magnetic fields, the asymptotic values of the oblique Hanle signals (in both contacts) are consistently reduced by a factor of $\cos^2(\theta)$ irrespective of the bias current and**

**temperature. These results are in good agreement with predictions of the spin precession and relaxation model for the electrical oblique Hanle effect. At high magnetic fields where the magnetization of CoFe is significantly tilted from the film plane to the magnetic field direction, we find that the observed angular dependence of voltage signals in the CoFe/MgO/Si and CoFe/MgO/Ge contacts are well explained by the OHE, considering the misalignment angle between the external magnetic field and the magnetization of CoFe.**

## I. INTRODUCTION

The electrical injection of spin-polarized electrons from a ferromagnet (FM) into a semiconductor (SC) and the subsequent detection of the resultant spin accumulation are the major building blocks of SC-based spintronics[1-5]. By engineering ferromagnetic tunnel contacts, the electrical injection and detection of the spin accumulation in various SC systems has been demonstrated up to room temperature (RT) through the Hanle effect[4,5].

The approach based on the Hanle effect[6,7], in which a magnetic field transverse to the spins suppresses the spin accumulation in a SC via spin precession and dephasing, provides an unambiguous means to establish the presence of spin accumulation in a SC. In particular, the oblique Hanle effect (i.e., the Hanle effect in an oblique magnetic field, OHE)[6-10] enables us to obtain additional information of spin dynamics and convincing proof of spin accumulation in the SC. The optical OHE using an optical detection technique (or the circular polarization of emitted light) in SC[8-11], has been intensely studied in spin light-emitting diodes (spin LEDs). However, the counterpart of the electrical OHE in SC still needs to be explored[6,7,12,13].

Here, we report the electrical investigation of the OHE in FM/oxide/SC contacts and their generic features using a three-terminal measurement scheme. The electrical OHE signals obtained in CoFe/MgO/Si and CoFe/MgO/Ge contacts show clearly different features depending on the spin relaxation time of the host SC. Notably, their asymptotic values (in both contacts) are consistently reduced by a factor of $\cos^2(\theta)$ in moderate magnetic fields at an oblique angle ($\theta$) to the spin direction irrespective of the bias current ($I$) and temperature ($T$). These results are highly consistent with predictions of the spin precession and relaxation model for the electrical OHE. The angular dependence of voltage signals obtained at high magnetic fields where the magnetization of CoFe is significantly tilted from the film plane to the magnetic field direction are also well explained by the same model, taking into account the misalignment angle between the external magnetic field and the magnetization of CoFe.

## II. EXPERIMENTAL DETAILS

### A. Device fabrication

Two types of CoFe(5 nm)/MgO(2 nm)/*n*-SC(001) tunnel contacts were prepared using a molecular beam epitaxy system. The first type is a highly ordered CoFe/MgO/Si contact in which the Si channel is heavily As-doped ($n_d \sim 2.5\times10^{19}$ cm$^{-3}$ at 300 K)[14], and the second is a single-crystalline CoFe/MgO/Ge contact in which the Ge channel consists of a heavily P-doped surface layer ($n_d \sim 10^{19}$ cm$^{-3}$ at 300 K) and a moderately Sb-doped substrate ($n_d \sim 10^{18}$ cm$^{-3}$ at 300 K)[14]. In order to measure the electrical OHE in the ferromagnetic tunnel contacts, we fabricated devices consisting of multiple CoFe/MgO/*n*-SC tunnel contacts (100×100 μm$^2$). Details of the sample preparation as well as the structural and electrical characterizations of the samples are available in the literature[14]. It should be noted that the dominant transport mechanism for both contacts (CoFe/MgO/Si and CoFe/MgO/Ge) is tunneling, as proven by the symmetric *I*-*V* curve and its weak *T*-dependence[14]. Furthermore, taking into account the similar roughness and magnetization (*M*) properties of both tunnel contacts, characterized by atomic force microscopy and vibrating sample magnetometry, respectively[15], it is likely that the magnitude of the local magnetostatic fields ($B_L^{ms}$) at the SC interface is similar in both contacts (note that $B_L^{ms}$ scales with the roughness of the FM interface and the magnetization of the FM)[16].

## B. Measurement scheme

Figures 1(a)-1(e) illustrate the electrical detection of the OHE in a FM/oxide/SC tunnel contact by means of the three-terminal measurement scheme, where a single ferromagnetic tunnel contact is used for electrical injection as well as for the detection of the spin accumulation in the SC[14-20]. The direction of the spin injection and the spin detection coincides with the direction of the magnetization (*M*) of the FM (see Figs. 1(d) and 1(e)).

The OHE signal depends on the magnitude and angle of the external magnetic fields. When an external magnetic field ($B^{ext}$) much smaller than $\mu_0 M_s$ of FM (where $\mu_0$ is the permeability of free space, and $M_s$ is the saturation magnetization) is applied at an oblique angle $\theta$ to the *x*-axis (Fig. 1(a)), the *M* of FM almost remains in-plane. The electrically injected spins in the SC under reverse bias ($I$<0) are precessed around the total magnetic field ($B^{tot}$) direction, given by the vector sum of $B_L^{ms}$ and $B^{ext}$. It is noteworthy that, even at zero $B^{ext}$, the injected spins are initially precessed and dephased by $B_L^{ms}$ having random directions at the SC interface[15,16].

At intermediate values of $B^{ext}$ ($\mu_0 M_s$ >>$B^{ext}$>> $B_L^{ms}$; Fig. 1(b)), the spins are precessed around $B^{ext}$ and the average spins are saturated along the oblique $B^{ext}$, resulting in an asymptotic value of spin signal depending on the angle $\theta$ (Fig. 1(d)). The resultant spin signal ($S$) in SC is reduced to $S_{0x}\cos(\theta)$. Because the same ferromagnetic tunnel contact is used to detect the spin signal, the magnitude of spin signal ($S_x$) at the detector is further reduced to $S_{0x}\cos^2(\theta)$.

When a very large $B^{ext}$ exceeding $\mu_0 M_s$ of the FM is applied (Figs. 1(c) and 1(e)), the $M$ of FM is significantly tilted from the film plane to the $B^{ext}$ direction with the tilting angle ($\varphi$). The misalignment angle ($\theta$-$\varphi$) between $B^{ext}$ and $M$ is determined by minimizing the total magnetic energy ($E_{tot}$) of the FM layer, consisting of the Zeeman energy and the demagnetization energy (or shape anisotropy energy).

$$E_{tot} = -M_s B_{ext}\cos(\theta-\phi) + \frac{1}{2}\mu_0 M_s^2 \sin^2(\phi), \qquad (1)$$

$$\varphi = \theta - \arctan\left( \operatorname{sgn}(\theta)\sqrt{\left(\frac{\cos(2\theta) + B_{ext}/\mu_0 M_s}{\sin(2\theta)}\right)^2 + 1} - \frac{\cos(2\theta) + B_{ext}/\mu_0 M_s}{\sin(2\theta)} \right). \qquad (2)$$

In this case, the misalignment angle ($\theta-\varphi$) can give rise to the OHE, resulting in the angular dependence of voltage signal proportional to $S_{0x'}\cos^2(\theta-\varphi)$ (Fig. 1(e)).

# III. RESULTS & DISCUSSION

## A. Model calculation of electrical oblique Hanle effect in ferromagnetic

**tunnel contacts**

To obtain insight into the generic features of the electrical OHE (in the ferromagnetic tunnel contact), we calculated the oblique Hanle curves for different spin lifetime values $(\tau_{sf})$ with a fixed value of $B_L^{ms}$ using the spin precession and relaxation model[16], including the initial spin precession and the dephasing due to the $B_L^{ms}$. Here we assumed the *M* of FM is in-plane ($B^{ext} << \mu_0 M_s$) for simplicity. The general case, where the *M* is tilted from the film plane (Fig. 1(c)), will be discussed in section C.

In the case of electrical spin injection $\vec{S}_i$ ($S_{0x}$,0 ,0) (see Fig. 1(c)), the $S_x$ component of the steady-state spin polarization $\vec{S}$ at the SC interface, which is parallel to the *M* direction of the FM detector, in $B^{tot}$ consisting of $B_L^{ms}$ and $B^{ext}$ is expressed as

$$S_x = S_{0x}\left\{\frac{\omega_x^2}{\omega_L^2} + \left(\frac{\omega_y^2+\omega_z^2}{\omega_L^2}\right)\left(\frac{1}{1+\left(\omega_L\tau_{sf}\right)^2}\right)\right\}, \qquad (3)$$

where $S_{0x}$ is the injected spin polarization without any magnetic field, $\omega_L$ $(=g\mu_B B_{tot}/\hbar)$ is the Larmor frequency, $\omega_L^2 = \omega_x^2 + \omega_y^2 + \omega_z^2$, and $\omega_i = \omega_i^{ext} + \omega_i^{ms}(x, y, z)$. Here, *g* is the Landé *g*-factor, $\mu_B$ is the Bohr magneton, $\hbar$ is the Planck constant divided by $2\pi$, and $\omega_i^{ms}(x, y, z)$ was set such that it had periodic spatial variation with $\omega_L^{ms}\cos(2\pi x/\lambda)$, where $\omega_L^{ms} \approx 3$ ns$^{-1}$ (or $1/\omega_L^{ms} \approx 0.33$ ns, corresponding to a $B_L^{ms}$ value of 0.3 kOe) and $\lambda$ =40 nm and where the spin polarization was averaged in space over a full period $\lambda$ for simplicity.

Two important features of the electrical OHE are obtained from the calculated

Hanle curves (normalized, Fig. 2(a)) at various angles ($0^0$ to $90^o$) with two different spin lifetimes ( $\tau_{sf}$ ) of 0.25 and 1.00 ns, respectively, at a fixed $1/\omega_L^{ms}$ value of 0.33 ns. For comparison, ideal OHE curves, where $B_L^{ms} \approx 0$ kOe (or $1/\omega_L^{ms} \approx \infty$ ns), are also shown (blue symbols; right panels of Fig. 2(a)). The first feature shows that the oblique Hanle line shapes are significantly dependent on $\tau_{sf}$ (at a fixed value of $1/\omega_L^{ms}$). For a large value of $\tau_{sf}$, the inverted OHE, indicative of the initial spin suppression due to $B_L^{ms}$ [15,16], becomes pronounced as the $\theta$ value approaches $0^o$; the width of the oblique Hanle curve at the angle $\theta$ of $90^o$ (red symbol) is remarkably broadened in comparison with the ideal Hanle curve (blue symbol) without $B_L^{ms}$, as the injected spins with a large value of $\tau_{sf}$ (strictly, $\tau_{sf} \geq 1/\omega_L^{ms}$) are precessed many times in $B_L^{ms}$ and randomized within their values of $\tau_{sf}$, resulting in the sizable suppression of the spin polarization and spin coherence, as discussed in the literature[15].

The second feature is that, in spite of the different features of the electrical OHE (depending on the values of $\tau_{sf}$ and $1/\omega_L^{ms}$), their asymptotic values at a high $B^{ext}$ are identical (see the top and bottom panels of Fig. 2(a)). For a more quantitative analysis, we plotted the asymptotic value of the OHE vs. $\theta$ for the two $\tau_{sf}$ values of 0.25 and 1.00 ns. As shown in Fig. 2(b), the asymptotic value of the electrical OHE depends only on the angle $\theta$, thus revealing the unique dependence on $\cos^2(\theta)$. This result is predicted by Eq. (3). When $B^{ext} >> B_L^{ms}$ (or $B^{ext} \approx B^{tot}$) and $\omega_L \tau_{sf} >> 1$, the normalized $S_x$ value is determined only by the ratio of the $B^{ext}$ component, $\left(B_x^{ext}\right)^2/\left(B^{ext}\right)^2$, which can be written in terms of the angle $\theta$, $\cos^2(\theta)$.

In addition, from the electrical OHE measurements, it is possible to extract the critical oblique angle ($\theta_c$), where the asymptotic value of the OHE signal coincides with the Hanle signal at zero $B^{ext}$ (see Fig. 2(a)), $\theta_c \equiv \arccos\left(\sqrt{S_x(B^{ext}=0)/S_{0x}}\right)$. This is a quantitative measure of the interfacial spin depolarization (ISD) due to the initial spin precession and dephasing by $B_L^{ms}$ [15,16].

## B. Experimental observation of the electrical oblique Hanle signals in CoFe/MgO/Si and CoFe/MgO/Ge contacts

We experimentally checked whether the electrical OHEs obtained in the CoFe/MgO/Si and CoFe/MgO/Ge contacts show the features consistent with the above model calculations. A constant bias current ($I$) is applied across the tunnel contact while the $V$ is measured as a function of applied $B^{ext}$ at a fixed angle $\theta$ (see Figs. 1(a) and 1(b)). It is noteworthy that the control experiments[4,5,17] using a nonmagnetic interfacial layer confirm that the observed Hanle signals in our system are genuine and arise from the spin accumulation.

Figures 3(a) and 3(b) show the obtained OHE signals ($|\Delta V_{OHE}|$) at various $\theta$ in the CoFe/MgO/Si and CoFe/MgO/Ge contacts, respectively, with applying $I$ of -0.5 mA (spin injection condition) at 300 and 5 K. At RT (the top panels of Figs. 3(a) and 3(b)), the CoFe/MgO/Si contact with an effective value of $\tau_{sf}$ ($\tau_{eff}$) of 167 ps, as extracted from the Lorentzian fit of the Hanle curve at an angle $\theta$ of 90$^o$ (note that the extracted $\tau_{eff}$ value should be considered as the lower bound for the $\tau_{sf}$ due to the artificial broadening of the Hanle curve caused by the $B_L^{ms}$)[15,16], shows a pronounced inverted Hanle signal at $\theta$ of 0$^o$ and 30$^o$. In contrast, the inverted Hanle effect is relatively weak for the CoFe/MgO/Ge contact having a $\tau_{eff}$ value of 104 ps. At a low $T$ of 5 K (bottom panels of Figs. 3(a) and 3(b)), the $\tau_{eff}$ values increase and the inverted OHEs at the $\theta$ values of 0$^o$ and 30$^o$ become larger; moreover, the increases of $\tau_{eff}$ and the inverted OHEs of the CoFe/MgO/Ge contact are more pronounced than those of the CoFe/MgO/Si contact, both of which are consistent with the findings in previous work[15]. Considering that the magnitude of the local $B_L^{ms}$ (at the SC interface) and the related $T$-dependence are not fundamentally different in both contacts, it is clear that the different features of the electrical OHE in the CoFe/MgO/Si and CoFe/MgO/Ge contacts are mainly ascribed to the different $\tau_{sf}$ values[15], as predicted by the model calculation.

Another important feature of the electrical OHE is the unique angular dependence of the asymptotic value of the oblique Hanle signal. In the model calculation (see Fig. 2(b)), it is expected that the asymptotic value at an intermediate value of $B^{ext}$ ($\mu_0 M_s >> B^{ext} >> B_L^{ms}$ (or $B^{ext} \approx B^{tot}$) and $\omega_L \tau_{sf} >> 1$) reveals the cos$^2$($\theta$) dependence on the angle $\theta$. To check this, we measured the asymptotic values of $|\Delta V_{OHE}|$ of the CoFe/MgO/Si and CoFe/MgO/Ge contacts as a function of the angle $\theta$ with oblique $B^{ext}$ values of 3 and 5 kOe, respectively.

An oblique $B^{ext}$ applied under the angle $\theta$ in the direction of the (in-plane) easy axis $M$ of the FM (x-axis, see Figs. 1(d) and (e)) will force $M$ to tilt out of plane by the amount of the tilting angle ($\varphi$). Using Eq. (2), we calculated the $\varphi$ variation with $\theta$ with the $B^{ext}$ of 3 kOe for CoFe/MgO/Si and 5 kOe for CoFe/MgO/Ge and the $\mu_0 M_s$ of CoFe (≈2.2 T). As depicted in Fig. 4(a), the $\varphi(\theta)$ values do not exceed 9$^{o}$ for CoFe/MgO/Si and 14$^{o}$ for CoFe/MgO/Ge contacts. Fig. 4 (b) shows that the calculated $\Delta V_{OHE,\,asym.value}(\theta)$ curves at the $B^{ext}$ of 3 and 5 kOe are very slightly deviated from the ideal case (no tilting; $\varphi$=0). From this, we can conclude that the tilting of $M$ with a $B^{ext}$ smaller than 5 kOe does not affect the major angular dependence of OHE signal proportional to $\cos^2(\theta)$ .

Figures 4(c) and 4(d) show the asymptotic value of $|\Delta V_{OHE}|$ (or $|\Delta V_{OHE,\,B=3\text{ or }5\text{ kOe}}|$) vs. $\theta$ obtained with the CoFe/MgO/Si and CoFe/MgO/Ge contacts, respectively, at the $I$ of -0.5 mA (spin injection condition) for different temperatures. These figures clearly show that the asymptotic values of the oblique Hanle signals in both contacts are consistently reduced by a factor of $\cos^2(\theta)$ (black lines) over a wide $T$ range (5-300 K). This result is in good agreement with the model prediction, in which the asymptotic value of the electrical OHE depends only on the angle $\theta$, revealing the unique $\cos^2(\theta)$ dependence.

From the electrical OHE measurements, we can extract the useful parameter of $\theta_c$, defined as $\theta_c \equiv \arccos\left(\sqrt{S_x(B_{ext}=0)/S_{0x}}\right)$, which is the quantitative measure of the ISD. The $\theta_c$ as a function of $T$ is plotted in Fig. 5. As $T$ decreases from 300 to 5 K, the value of $\theta_c$ increases gradually; the $\theta_c$ values vary from 52$^{o}$ and 36$^{o}$ (300 K) to 58$^{o}$ and 66$^{o}$ (5 K) for the CoFe/MgO/Si and CoFe/MgO/Ge contacts, respectively. Qualitatively the same behavior was observed in the $T$-dependence of $\theta_c$ obtained at the constant bias voltage ($V_{B=0}$) of -0.3 V (spin injection condition; not shown). This result is mainly ascribed to the enhanced ISD effect[15] owing to the increased $\tau_{sf}$ at a low $T$ (note that the $T$-dependence of $\omega_L^{ms}$ is weak because $B_L^{ms}(T) \propto (1-\alpha T^{3/2})$ with $\alpha = 3.2\times10^{-5} K^{-3/2}$ for the CoFe)[21,22].

## C. Angular dependence of voltage signals at high magnetic fields: anisotropic spin accumulation

Next we have investigated the angular dependence of voltage ($|\Delta V(\theta)|$) signals when the applied $B^{ext}$ is larger than $\mu_0 M_s$. In this case, as discussed previously, the $M$ is significantly tilted from the film plane (see Figs. 1(c) and 1(e)) to the $B^{ext}$ direction. The $|\Delta V(\theta)|$ signals (at high magnetic fields) were measured as a function of angle $\theta$ (from in-plane to out-of-plane) at the fixed $B^{ext}$ value of 50 kOe. The misalignment angle ($\theta-\varphi$) between the applied $B^{ext}$ and the $M$ of the CoFe can be determined by using Eq. (2). Figure 6(a) shows the calculated angle ($\theta-\varphi$) as a function of angle $\theta$ with the $B^{ext}$ of 50 kOe and the $\mu_0 M_s$ of CoFe (~2.2 T). This misalignment angle ($\theta-\varphi$) gives rise to the OHE, resulting in the anisotropic spin accumulation ($\Delta\mu_{\mathrm{OHE}}(\theta)$) with the $\cos^2(\theta-\varphi)$ angular dependence, as depicted in Fig. 6(b). In this figure, we assumed that the angular dependences of tunnel spin polarization and spin lifetime values[23] are negligible for simplicity. The important consequence obtained from the $\Delta\mu_{\mathrm{OHE}}\,(\mathrm{normalized})$-$\theta$ curve in Fig. 6(b) is that the spin accumulation signal is reduced depending on the field angle ($\theta$). The minima occurs at $\theta$ angles of ~-58$^{o}$ and ~58$^{o}$ and the ratio of its suppressed portion to the full spin accumulation, where $\theta=\varphi$, is ~0.048.

Figure 7 shows the $|\Delta V(\theta)|$ signals of the CoFe/MgO/Si and CoFe/MgO/Ge

contacts measured at different magnetic field ranges with various reverse bias currents ($I$<0; spin injection condition) at 5 K. Figures 7 (a) and 7(b), respectively, are obtained with applying 3 kOe for the CoFe/MgO/Si contact and 5 kOe for the CoFe/MgO/Ge contact. In this case, the $\left|\Delta V_{B=3\text{ or }5\text{ kOe}}\right|$ signals in both contacts clearly reveal $\cos^2(\theta)$ dependence (black lines), irrespective of $I$, having the twofold symmetry with the peak at an angle $\theta$ of $0^o$ as well as valleys at $\theta$ angles of $-90^o$ and $+90^o$. The magnitudes of the MR values, defined as $\left(\Delta V / V_{\theta=90^o}\Big|_{B=3\text{ or}5\text{ kOe}}\right)\times 100$ %, in both contacts are about 1 % at 5 K.

Figures 7(c) and 7(d), respectively, are obtained with applying 50 kOe for the CoFe/MgO/Si and CoFe/MgO/Ge contacts. The $\left|\Delta V_{B=50\text{ kOe}}\right|$ signals in Figs. 7(c) and 7(d) show distinctly different characteristics from the $\left|\Delta V_{B=3\text{ or }5\text{ kOe}}\right|$ in Figs. 7(a) and 7(b). Basically, the $\left|\Delta V_{B=50\text{ kOe}}\right|$ signals of both contacts show the distorted fourfold symmetry with valleys at $\theta$ angles of $-60^o$ and $60^o$ and peaks at $\theta$ angles of $-90^o$, $0^o$, and $90^o$. The MR values, defined as $\left(\Delta V / V_{\theta=60^o}\Big|_{B=50\text{ kOe}}\right)\times 100$ %, of ~0.05 (CoFe/MgO/Si) and ~0.10 % (CoFe/MgO/Ge) at 5 K are roughly one order of magnitude smaller than the MR measured at moderate magnetic fields (3 or 5 kOe). This result is quite in good agreement with the $\cos^2(\theta-\varphi)$ angular dependence as depicted in Fig. 6(b). The fair $\cos^2(\theta-\varphi)$ fits (black lines in Figs. 7(c) and 7(d)) to the experimental data of $\left|\Delta V_{B=50\text{ kOe}}\right|$ in the CoFe/MgO/Si and CoFe/MgO/Ge contacts implies that the angular dependence of voltage signals at high magnetic fields in both contacts are consequence of the anisotropic spin accumulation (ASA) due to the OHE caused by the misalignment between the $B^{ext}$ and the $M$ of CoFe (see Fig. 1(e)). The deviation of the CoFe/MgO/Ge data from the $\cos^2(\theta-\varphi)$ fit when the $\theta$ value is close to $0^0$ (black arrows in Fig. 5(d)) is possibly attributed to the Lorenz MR (LMR) of the Ge substrate because Ge has a relatively high mobility and the resultant LMR is quadratic in terms of the mobility and the transverse $B$ to the current flow.

To confirm that the ASA is a main origin of the $|\Delta V(\theta)|$ signals obtained at 50 kOe in the CoFe/MgO/Si and CoFe/MgO/Ge contacts, we have compared the bias dependence of $|\Delta V_{OHE,\,B=50\,\mathrm{kOe}}|$ with that of the full spin accumulation $|\Delta V_{spin}|$, which is given by the sum of normal and inverted Hanle signal. The $|\Delta V_{OHE,\,B=50\,\mathrm{kOe}}|$ is defined as $\left|\Delta V_{\theta=90^{\mathrm{o}}} - \Delta V_{\theta=60^{\mathrm{o}}}\right|_{B=50\,\mathrm{kOe}}$ (see Figs. 7 (c) and (d)) and the $|\Delta V_{spin}|$ is defined as $\left|\Delta V_{\theta=0^{\mathrm{o}}} - \Delta V_{\theta=90^{\mathrm{o}}}\right|_{B=3\text{ or}5\text{ kOe}}$ (see Figs. 7(a) and (b)). Figures 8(a) and 8(b) respectively shows the measured $|\Delta V_{spin}|$ and $|\Delta V_{OHE,\,B=50\,\mathrm{kOe}}|$ values of the CoFe/MgO/Si and CoFe/MgO/Ge contacts with various reverse bias currents ($I$<0; spin injection condition) at 5 K. In these figures, one can clearly see that the variation of $|\Delta V_{OHE,\,B=50\,\mathrm{kOe}}|$ values with the bias current is similar to that of the $|\Delta V_{spin}|$. More importantly, $|\Delta V_{OHE,\,B=50\,\mathrm{kOe}} / \Delta V_{\mathrm{spin}}|$, the ratio of reduced spin signal due to the misalignment-induced OHE, in Fig. 8(c) is ~0.060 (0.065) for the CoFe/MgO/Si (CoFe/MgO/Ge) contact and remains almost constant with varying the $I$. This is very close to the expected value of ~0.048 (black circles) from the model calculation. These results strongly indicate that the observed $|\Delta V(\theta)|$ signals at high magnetic fields in CoFe/MgO/Si and CoFe/MgO/Ge contacts (under spin injection condition) mainly originate from the ASA (owing to the misalignment-induced OHE).

It should be finally mentioned that, according to the recent report[23], the out-of-plane tunneling anisotropy (TA) signals in FM/$Al_2O_3$/Si contacts come from the two different sources of the regular tunneling anisotropic magnetoresistance (TAMR)[24-29] and the ASA[23]. It is known that the regular TAMR[24-29] is associated with the anisotropic density of states at FM/insulator tunnel interface, the spin-orbit coupling, and the spin-dependent scattering; the ASA[23] is attributed to the Hanle spin precession (i.e., OHE) due to the misalignment, the anisotropic tunnel spin polarization at FM/insulator tunnel interface[30,31], and the anisotropic spin relaxation in the SC. Among them, it is found here that the ASA (owing to the misalignment-induced OHE) dominantly contributes to the $|\Delta V(\theta)|$ signals at high magnetic fields in the crystalline CoFe/MgO/Si and CoFe/MgO/Ge contacts (under spin injection condition).

## IV. CONCLUSIONS

In conclusion, we investigated the electrical OHE in CoFe/MgO/Si and CoFe/MgO/Ge contacts and their generic features using a three-terminal measurement scheme. The electrical OHE signals obtained in the CoFe/MgO/Si and CoFe/MgO/Ge contacts show clearly different line shapes depending on the spin lifetime of the host SC. Importantly, irrespective of the bias current and temperature, the asymptotic values of the OHE in both contacts reveal the universal angular dependence with $\cos^2(\theta)$ variation. These results are highly consistent with the predictions of spin precession and relaxation model for the electrical OHE. The angular dependence of voltage signals observed at high magnetic fields where the magnetization of CoFe is significantly tilted from the film plane to the magnetic field direction, are also well explained by the OHE, taking into account the misalignment angle between the external magnetic field and the magnetization of CoFe.

## ACKNOWLEDGMENTS

This work was supported by the DGIST R&D Program of the Ministry of Education, Science and Technology of Korea (11-IT-01); by the KIST institutional program (2E22732 and 2V02720) and by the Pioneer Research Center Program (2011-0027905); and by the KBSI grant no.T32517 for S-YP.

## REFERENCES

[1]A. Fert, *Rev. Mod. Phys.* **80**, 1517 (2008).

[2]I. Žutić, J. Fabian, and S. Das Sarma, *Rev. Mod. Phys.* **76**, 323 (2004).

[3]S. A. Wolf, D. D. Awschalom, R. A. Buhrman, J. M. Daughton, S. von Molnár, M. L. Roukes, A. Y. Chtchelkanova, and D. M. Treger, *Science* **294**, 1488 (2001).

[4]R. Jansen, *Nature Mater.* **11**, 400 (2012).

[5]R. Jansen, S. P. Dash, S. Sharma, and B. C. Min, *Semicond. Sci. Technol.* **27**, 083001 (2012).

[6]M. Johnson and S. H. Silsbee, *Phys. Rev. Lett.* **55**, 1790 (1985).

[7]M. Johnson and S. H. Silsbee, *Phys. Rev. B* **37**, 5326 (1988).

[8]M. I. D'yakonov, V. I. Perel', V. L. Berkovits, and V. I. Safarov, *Sov. Phys. JETP* **40**, 950 (1975).

[9]F. Meier and B. P. Zakharchenya, *Optical Orientation* (North-Holland, Amsterdam, 1984).

[10]V. F. Motsnyi, P. Van Dorpe, W. Van Roy, E. Goovaerts, V. I. Safarov, G. Borghs, and J. De Boeck, *Phys. Rev. B* **68**, 245319 (2003).

[11]V. F. Motsnyi, V. I. Safarov, J. De Boeck, J. Das, W. Van Roy, E. Goovaerts, and G. Borghs, *Appl. Phys. Lett.* **81**, 265 (2002).

[12]J. Li, B. Huang, and I. Appelbaum, *Appl. Phys. Lett.* **92**, 142507 (2008).

[13]C. Awo-Affouda, O. M. J. van ’t Erve, G. Kioseoglou, A. T. Hanbicki, M. Holub, C. H. Li, and B. T. Jonker, *Appl. Phys. Lett.* **94**, 102511 (2009).

[14]K. R. Jeon, C. Y. Park, and S. C. Shin, *Cryst. Growth Des.* **10**, 1346 (2010); K. R. Jeon, B. C. Min, Y. H. Park, H. S. Lee, C. Y. Park, Y. H. Jo, and S. C. Shin, *Appl. Phys. Lett.* **99**, 162106 (2011); K. R. Jeon, B. C. Min, I. J. Shin, C. Y. Park, H. S. Lee, Y. H. Jo, and S. C. Shin, *Appl. Phys. Lett.* **98**, 262102 (2011).

[15]K. R. Jeon, B. C. Min, Y. H. Park, Y. H. Jo, S. Y. Park, C. Y. Park, and S. C. Shin, *Appl. Phys. Lett.* **101**, 022401 (2012).

[16]S. P. Dash, S. Sharma, J. C. Le Breton, H. Jaffrès, J. Peiro, J. M. George, A. Lemaitre, and R. Jansen, *Phys. Rev. B* **84**, 054410 (2011).

[17]K. R. Jeon, B. C. Min, Y. H. Jo, H. S. Lee, I. J. Shin, C. Y. Park, S. Y. Park, and S. C. Shin, Phys. Rev. B **84**, 165315 (2011).

[18]X. Lou, C. Adelmann, M. Furis, S. A. Crooker, C. J. Palmstrøm, and P. A. Crowell, *Phys. Rev. Lett.* **96**, 176603 (2006).

[19]S. P. Dash, S. Sharma, R. S. Patel, M. P. de Jong, and R. Jansen, *Nature* **462**, 491 (2009).

[20]C. H. Li, O. M. J. van 't Erve, and B. T. Jonker, *Nat. Commun.* **2**, 245 (2011).

[21]W. Kipferl, M. Dumm, M. Rahm, and G. Bayreuther, *J. Appl. Phys.* **93**, 7601 (2003).

[22]W. Wang, H. Sukegawa, and K. Inomata, *Phys. Rev. B* **82**, 092402 (2010).

[23]S. Sharma, S. P. Dash, H. Saito, S. Yuasa, B. J. van Wees, and R. Jansen, *Phys. Rev. B* **86**, 165308 (2012).

[24]C. Gould, C. Ruster, T. Jungwirth, E. Girgis, G. M. Schott, R. Giraud, K. Brunner, G. Schmidt, L. W. Molenkamp, *Phys. Rev. Lett*. **93**, 117203 (2004).

[25]H. Saito, S. Yuasa, and K. Ando, *Phys. Rev. Lett.* **95**, 086604 (2005).

[26]L. Gao, X. Jiang, S. H. Yang, J. D. Burton, E. Y. Tsymbal, and S. P. P. Parkin, *Phys. Rev. Lett.* **99**, 226602 (2007).

[27]J. Moser, A. Matos-Abiague, D. Schuh, W. Wegscheider, J. Fabian, and D. Weiss, *Phys. Rev. Lett.* **99**, 056601 (2007).

[28]B. G. Park, J. Wunderlich, D. A. Williams, S. J. Joo, K. Y. Jung, K. H. Shin, K. Olejník, A. B. Shick, and T. Jungwirth, *Phys. Rev. Lett.* **100**, 087204 (2008).

[29]T. Akiho, T. Uemura, M. Harada, K. –i. Matsuda, and M. Yamamoto, *Appl. Phys. Lett.* **98**, 232109 (2010).

[30]A. Einwanger, M. Ciorga, U. Wurstbauer, D. Schuh, W. Wegscheider, and D. Weiss, *Appl. Phys. Lett.* **95**, 152101 (2009).

[31]M. Ciorga, A. Einwanger, U. Wurstbauer, D. Schuh, W. Wegscheider, and D. Weiss, *Physica E* **42**, 2673 (2010).

## FIGURE CAPTIONS

FIG. 1. (Color online) (a)-(e) Schematic illustration of the electrical detection of the OHE in a FM/oxide/SC tunnel contact for different external magnetic field ($B^{ext}$) ranges using the three-terminal measurement scheme.

FIG. 2. (Color online) (a) Calculated Hanle curves for various oblique angles ($0^0$ to $90^o$) with two different $\tau_{sf}$ values of 0.25 and 1.00 ns, respectively, at a fixed $1/\omega_L^{ms}$ value

of 0.33 ns. For comparison, the ideal OHE curves, where $B_L^{ms} \approx 0$ kOe (or $1/\omega_L^{ms} \approx \infty$ ns), are also shown (blue symbols). (b) Asymptotic values of the OHE vs. $\theta$ for the two $\tau_{sf}$ values of 0.25 and 1.00 ns.

FIG. 3. (Color online) Obtained OHE signals ( $|\Delta V_{OHE}|$) in the (a) CoFe/MgO/Si and (b) CoFe/MgO/Ge contacts, respectively, at the bias current (*I*) of -0.5 mA (spin injection condition) with various $\theta$ values for 300 and 5 K.

FIG. 4. (Color online) (a) The tilting angle ($\varphi$) variation with $\theta$ at the $B^{ext}$ values of 3 (for CoFe/MgO/Si) and 5 kOe (for CoFe/MgO/Ge). (b) Calculated $\Delta V_{OHE,\, asym.value}(\theta)$ curves for the $B^{ext}$ values of 3 (for CoFe/MgO/Si) and 5 kOe (for CoFe/MgO/Ge). For comparison, the ideal $\Delta V_{OHE,\, asym.value}(\theta)$ curve, where no tilting ($\varphi$=0), is also shown (black symbol). Asymptotic value of $|\Delta V_{OHE}|$ (or $|\Delta V_{OHE, B=3 \text{ or} 5 \text{ kOe}}|$) vs. $\theta$ obtained in (c) CoFe/MgO/Si and (d) CoFe/MgO/Ge contacts at the *I* value of -0.5 mA (the spin injection condition) at different temperatures.

FIG. 5. (Color online) Measured critical oblique angles ($\theta_c$) in the (a) CoFe/MgO/Si and (b) CoFe/MgO/Ge contacts as a function of the temperature (*T*).

FIG. 6. (Color online) (a) Calculated misalignment angle ($\theta-\varphi$) and (b) corresponding $\Delta\mu_{\text{OHE}}$ (normalized) as a function of $B^{ext}$ angle $\theta$ at the $B^{ext}$ and $\mu_0 M_s$ values of 50 kOe and ~2.2 T, respectively.

FIG. 7. (Color online) (a)/(b) $\left|\Delta V_{B=3\mathrm{or}5\,\mathrm{kOe}}\right|$ and (c)/(d) $\left|\Delta V_{B=50\,\mathrm{kOe}}\right|$ signals vs. $\theta$ curves for the CoFe/MgO/Si contact ((a), (c)) and the CoFe/MgO/Ge contact ((b), (d)) with various reverse bias current ($I$<0; the spin injection condition) at 5 K. Black lines in Figs. (a) and (b) are $\cos^2(\theta)$ fits. Black lines in Figs. (c) and (d) are $\cos^2(\theta-\varphi)$ fits.

FIG. 8. (Color online) (a) Measured $\left|\Delta V_{spin}\right|$, (b) $\left|\Delta V_{OHE,\,B=50\,\mathrm{kOe}}\right|$, and (c) $\left|\Delta V_{OHE,\,B=50\,\mathrm{kOe}}/\Delta V_{\mathrm{spin}}\right|$ values for the CoFe/MgO/Si and CoFe/MgO/Ge contacts with various reverse bias currents ($I$<0; spin injection condition) at 5 K. The black circles represent the expected value from the model calculation.

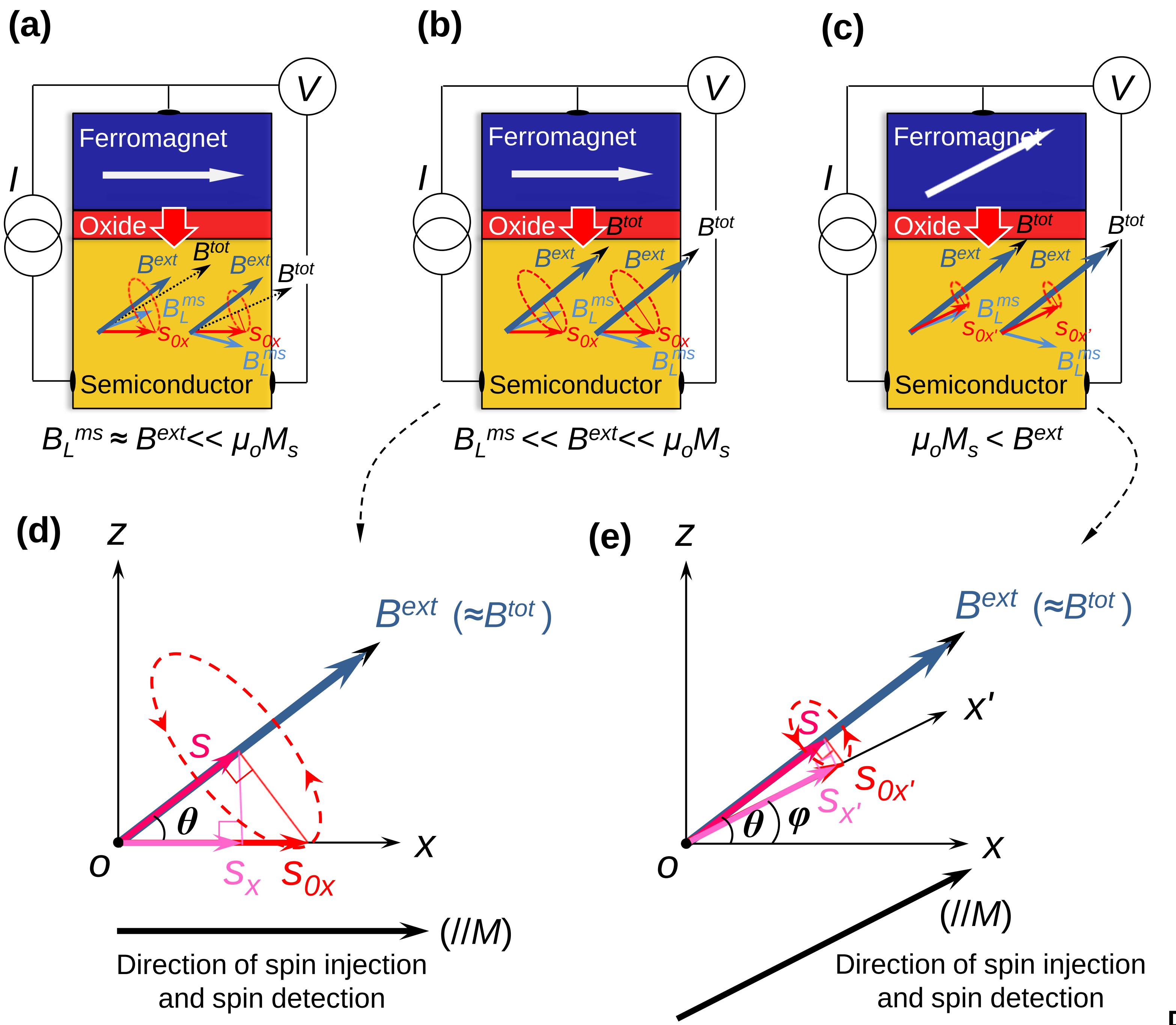

Fig. 1

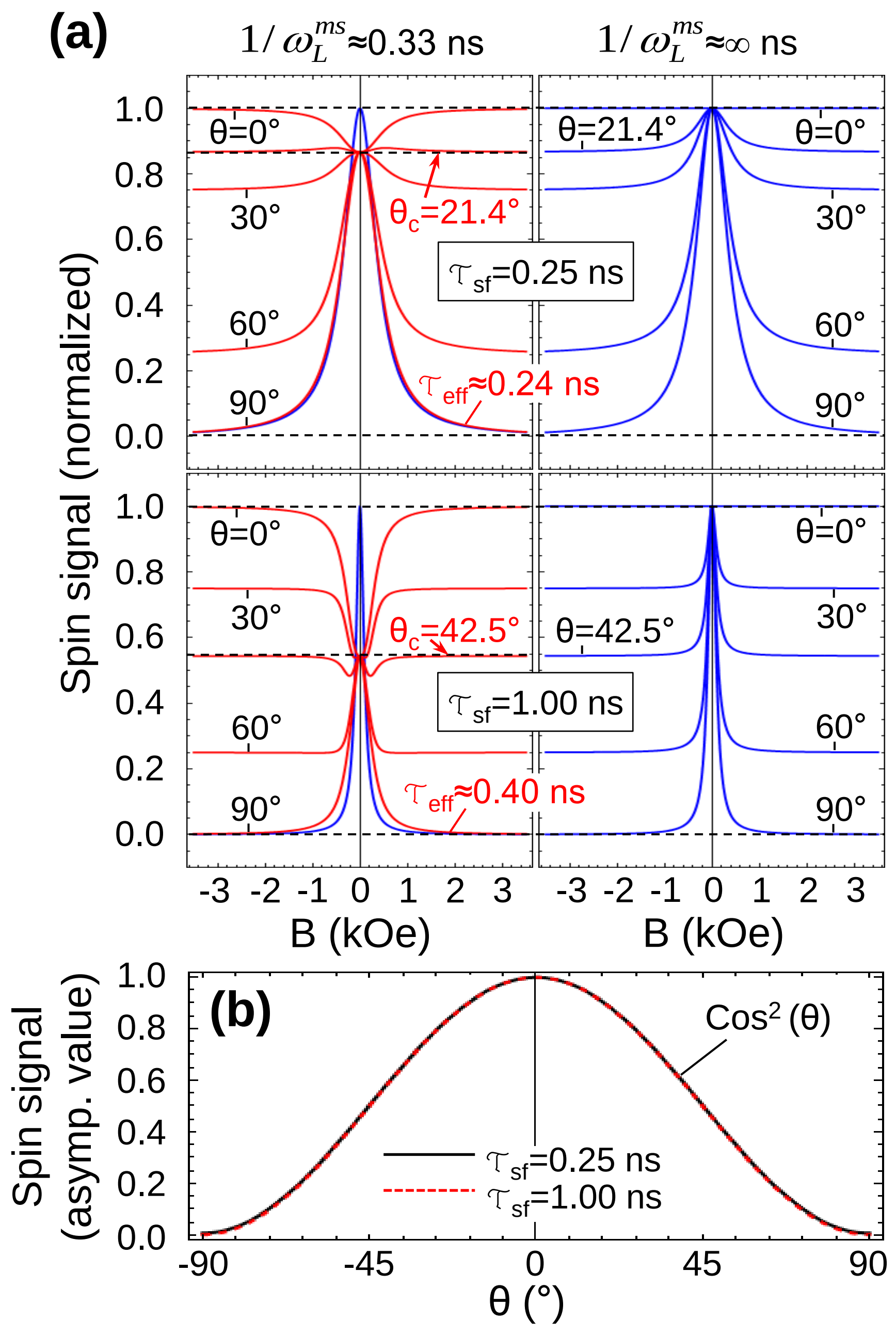

Fig. 2

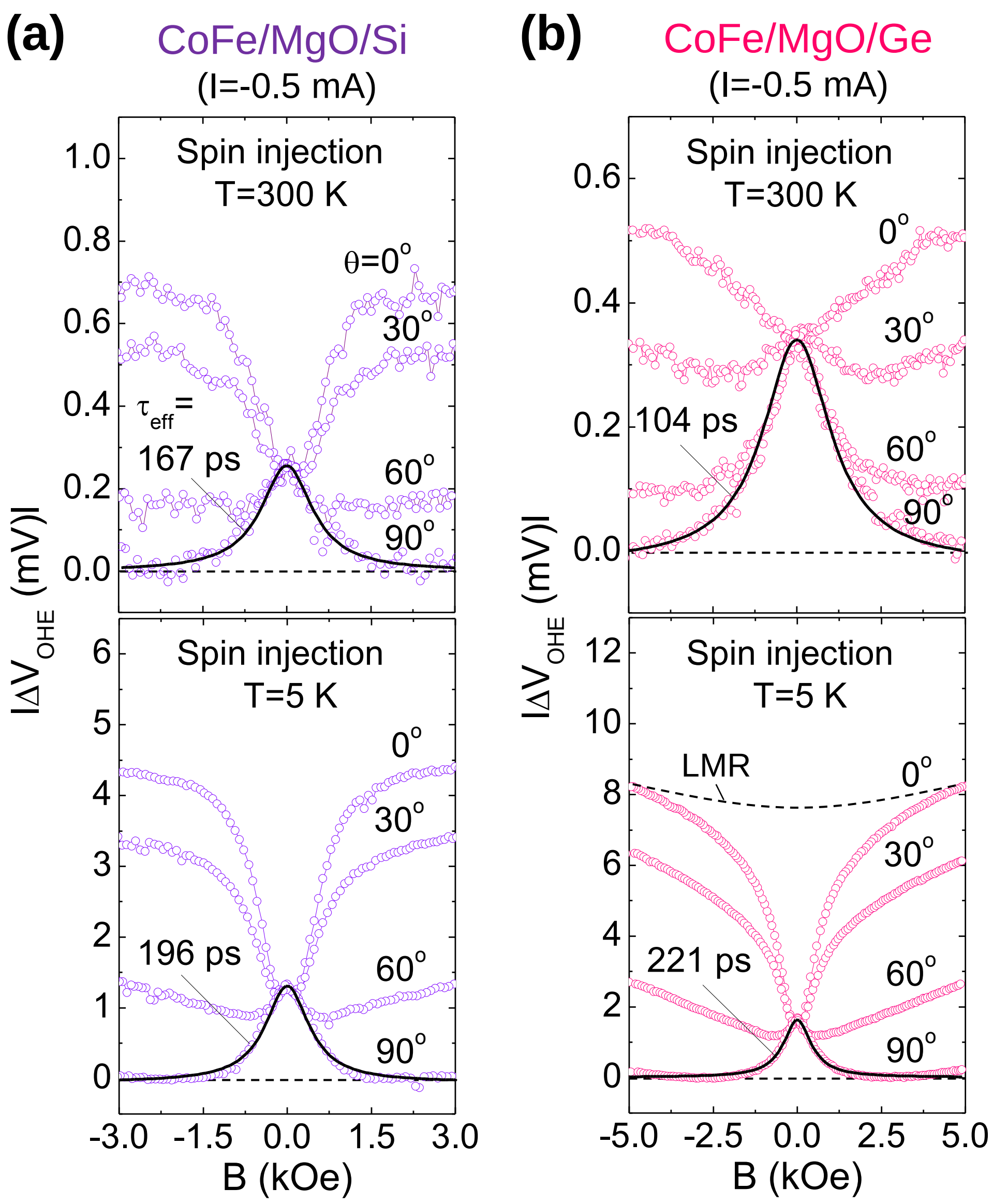
(a)
CoFe/MgO/Si
(I=-0.5 mA)
Spin injection
T=300 K
θ=0°
30°
60°
90°
τ_eff=
167 ps
1.0
0.8
0.6
0.4
0.2
0.0
Spin injection
T=5 K
0°
30°
60°
90°
196 ps
6
5
4
3
2
1
0
-3.0
-1.5
0.0
1.5
3.0
B (kOe)
|ΔV_OHE (mV)|
(b)
CoFe/MgO/Ge
(I=-0.5 mA)
Spin injection
T=300 K
0°
30°
60°
90°
104 ps
0.6
0.4
0.2
0.0
Spin injection
T=5 K
LMR
0°
30°
60°
90°
221 ps
12
10
8
6
4
2
0
-5.0
-2.5
0.0
2.5
5.0
B (kOe)
|ΔV_OHE (mV)|

Fig. 3

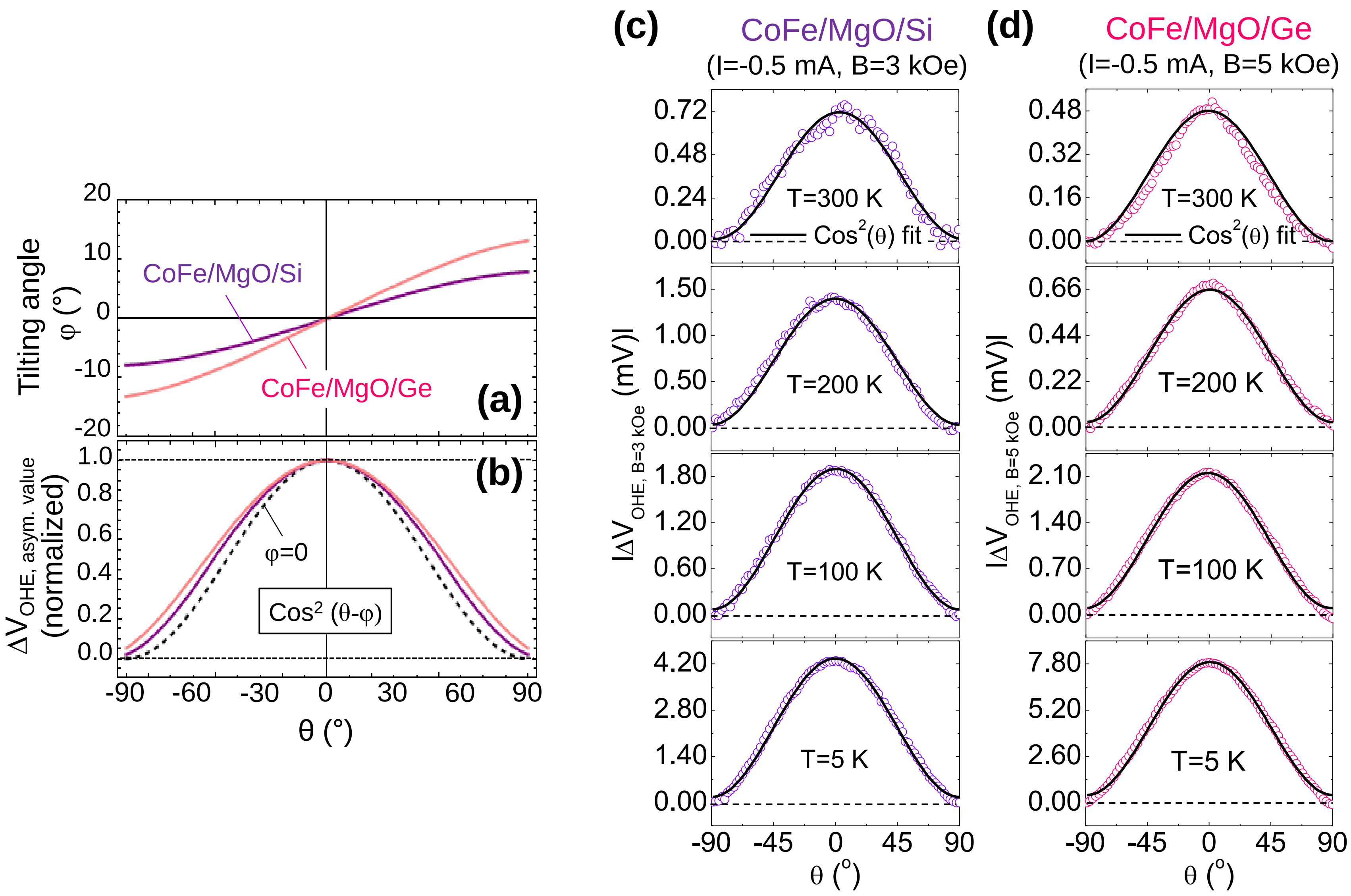

(a)
Tilting angle φ (°)
CoFe/MgO/Si
CoFe/MgO/Ge
20
10
0
-10
-20
(b)
ΔV OHE, asym. value (normalized)
φ=0
Cos² (θ-φ)
1.0
0.8
0.6
0.4
0.2
0.0
-90
-60
-30
0
30
60
90
θ (°)
(c)
CoFe/MgO/Si
(I=-0.5 mA, B=3 kOe)
|ΔV OHE, B=3 kOe (mV)|
T=300 K
Cos²(θ) fit
T=200 K
T=100 K
T=5 K
0.72
0.48
0.24
0.00
1.50
1.00
0.50
0.00
1.80
1.20
0.60
0.00
4.20
2.80
1.40
0.00
-90
-45
0
45
90
θ (°)
(d)
CoFe/MgO/Ge
(I=-0.5 mA, B=5 kOe)
|ΔV OHE, B=5 kOe (mV)|
T=300 K
Cos²(θ) fit
T=200 K
T=100 K
T=5 K
0.48
0.32
0.16
0.00
0.66
0.44
0.22
0.00
2.10
1.40
0.70
0.00
7.80
5.20
2.60
0.00
-90
-45
0
45
90
θ (°)

Fig. 4

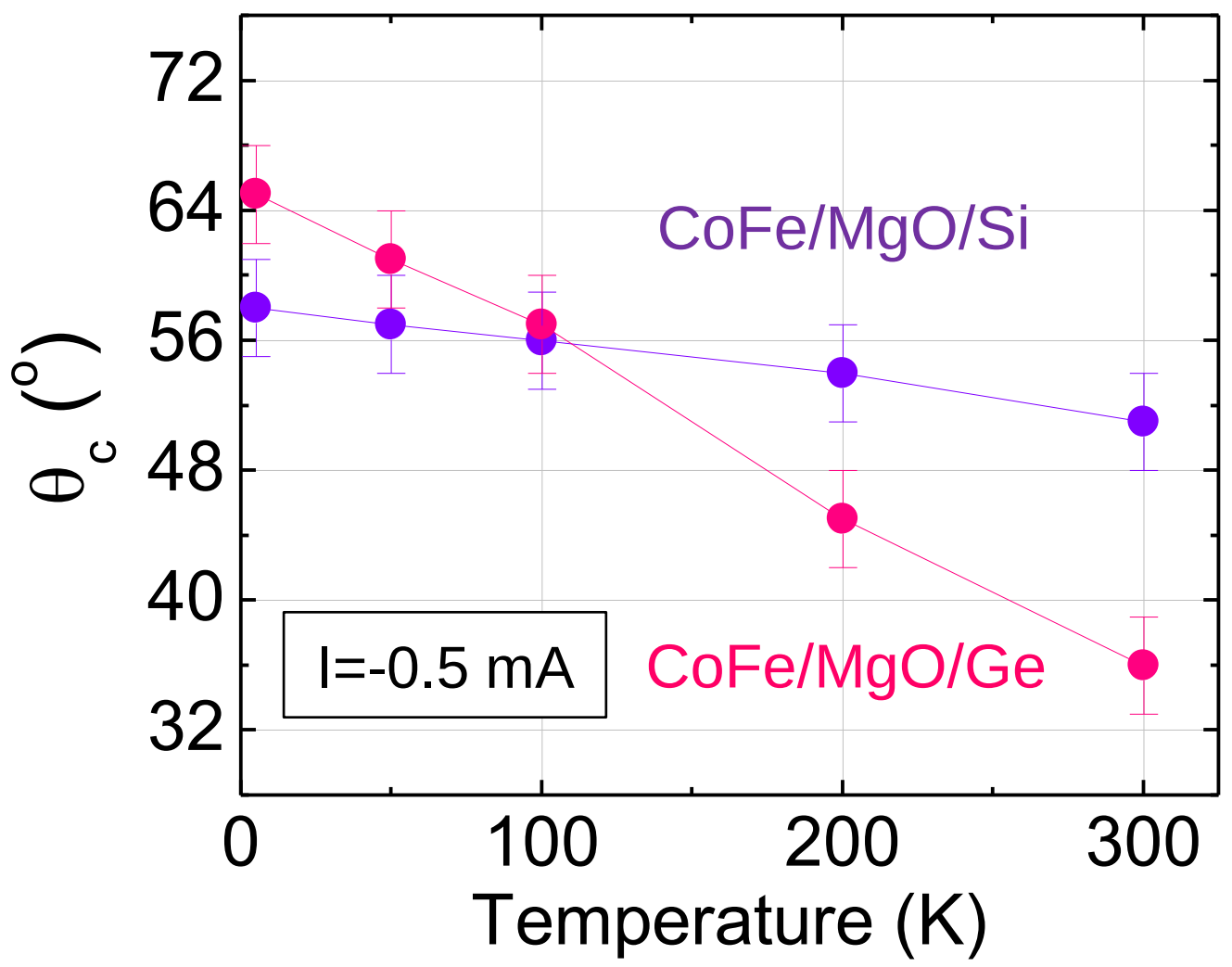

Fig. 5

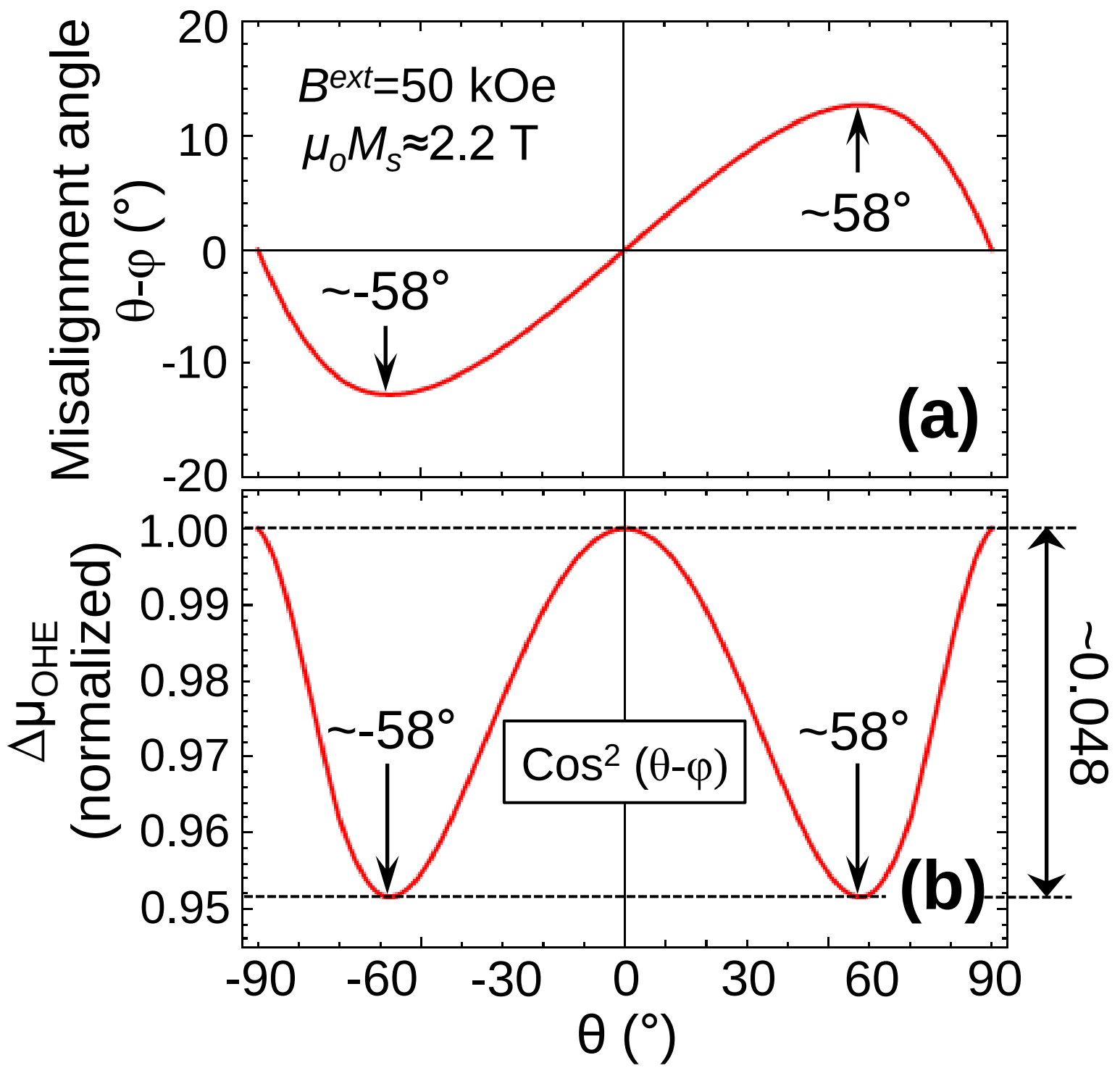

Fig. 6

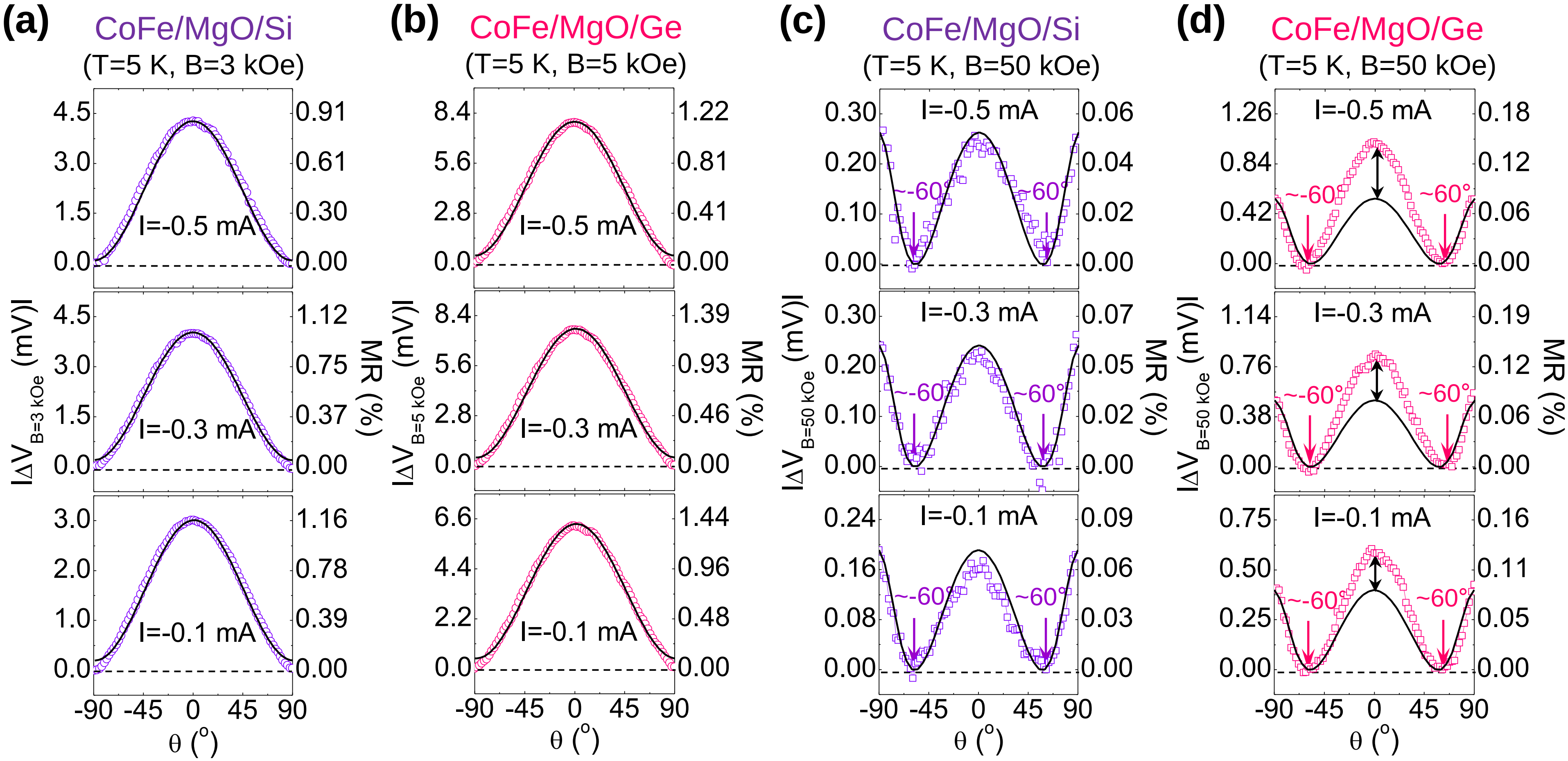
(a)
CoFe/MgO/Si
(T=5 K, B=3 kOe)
I=-0.5 mA
I=-0.3 mA
I=-0.1 mA
$|\Delta V_{B=3\ kOe}|$ (mV)
MR (%)
(b)
CoFe/MgO/Ge
(T=5 K, B=5 kOe)
I=-0.5 mA
I=-0.3 mA
I=-0.1 mA
$|\Delta V_{B=5\ kOe}|$ (mV)
MR (%)
(c)
CoFe/MgO/Si
(T=5 K, B=50 kOe)
I=-0.5 mA
I=-0.3 mA
I=-0.1 mA
~-60°
~60°
$|\Delta V_{B=50\ kOe}|$ (mV)
MR (%)
(d)
CoFe/MgO/Ge
(T=5 K, B=50 kOe)
I=-0.5 mA
I=-0.3 mA
I=-0.1 mA
~-60°
~60°
$|\Delta V_{B=50\ kOe}|$ (mV)
MR (%)
θ (°)

Fig. 7

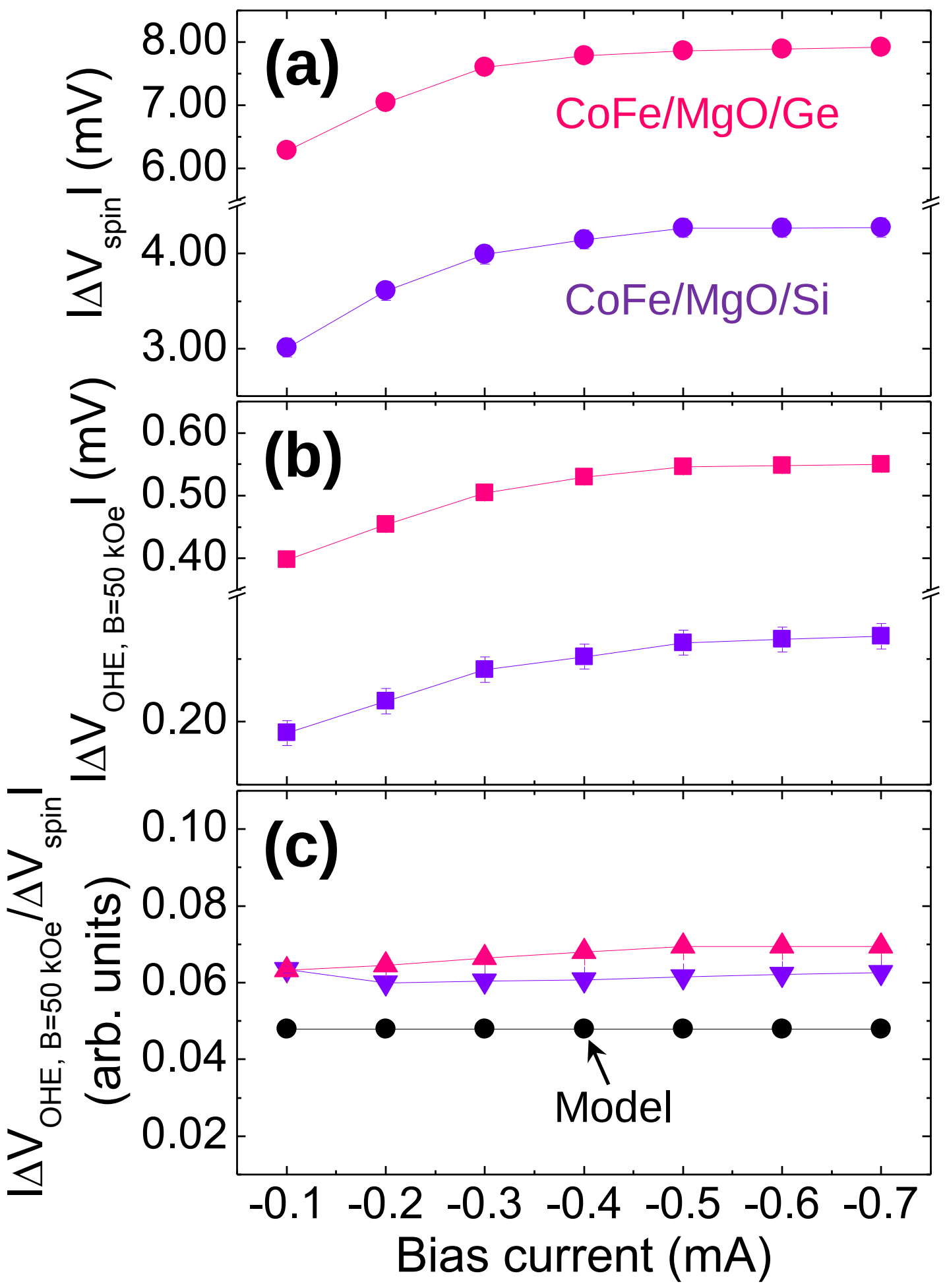
(a)
CoFe/MgO/Ge
CoFe/MgO/Si
|ΔV_spin| (mV)
8.00
7.00
6.00
4.00
3.00
(b)
|ΔV_OHE, B=50 kOe| (mV)
0.60
0.50
0.40
0.20
(c)
|ΔV_OHE, B=50 kOe/ΔV_spin|
(arb. units)
0.10
0.08
0.06
0.04
0.02
Model
-0.1
-0.2
-0.3
-0.4
-0.5
-0.6
-0.7
Bias current (mA)

Fig. 8